\RequirePackage{fix-cm}
\documentclass[twocolumn]{svjour3}          
\smartqed  
\usepackage{graphicx}
\usepackage{mathptmx}      
\usepackage{latexsym}
\usepackage[square,numbers]{natbib}
\usepackage{amsmath}
\usepackage{hyperref}
\DeclareMathOperator*{\argmin}{argmin}
\journalname{Neural Computing and Applications}

\begin{document}

\title{Applying Visual Domain Style Transfer and Texture Synthesis Techniques to Audio - Insights and Challenges%
\thanks{This research was supported by an NVIDIA Corporation Academic Programs GPU grant.}
}


\author{Muhammad Huzaifah         \and
        Lonce Wyse 
}


\institute{Muhammad Huzaifah bin Md Shahrin\at
             National University of Singapore, NUS Graduate School for Integrative Sciences and Engineering \\
              ORCID: 0000-0002-7188-3600\\
              \email{E0029863@u.nus.edu}           
           \and
           Lonce Wyse \at
              National University of Singapore, Communications and New Media Department\\
              ORCID: 0000-0002-9200-1048\\
              \email{cnmwll@nus.edu.sg}
}

\date{This is a post-peer-review, pre-copyedit version of an article published in Neural Computing and Applications. \\
 The final authenticated version is available online at: http://dx.doi.org/10.1007/s00521-019-04053-8} 

\maketitle

\begin{abstract}
Style transfer is a technique for combining two images based on the activations and feature statistics in a deep learning neural network architecture. This paper studies the analogous task in the audio domain and takes a critical look at the problems that arise when adapting the original vision-based framework to handle spectrogram representations. We conclude that CNN architectures with features based on 2D representations and convolutions are better suited for visual images than for time-frequency representations of audio. Despite the awkward fit, experiments show that the Gram matrix determined “style” for audio is more closely aligned with timbral signatures without temporal structure whereas network layer activity determining audio “content” seems to capture more of the pitch and rhythmic structures. We shed insight on several reasons for the domain differences with illustrative examples. We motivate the use of several types of one-dimensional CNNs that generate results that are better aligned with intuitive notions of audio texture than those based on existing architectures built for images. These ideas also prompt an exploration of audio texture synthesis with architectural variants for extensions to infinite textures, multi-textures, parametric control of receptive fields and the constant-Q transform as an alternative frequency scaling for the spectrogram.   
\end{abstract}

\section{Introduction and related work}
\label{intro}

If articulating the intuitive distinction between style and content for images is difficult, it is even more so for sound, in particular non-speech sound. The recent use of neural network-derived statistics to describe such perceptual qualities for texture synthesis \cite{gatys2015texture} and style transfer \cite{gatys2015neural} offers a fresh computational outlook on the subject. This paper examines several issues with these image-based techniques when adapting them for the audio domain. 

Belonging to a class of problems associated with generative applications, style transfer as a concept within deep learning was popularized with the pioneering work of Gatys et al. \cite{gatys2015neural,gatys2015texture}. Their core approach leverages the convolutional neural network (CNN) to extract high-level feature representations that correspond to certain perceptual dimensions of an image. In texture synthesis \cite{gatys2015texture} image statistics were extracted from a single reference texture with the objective of generating further examples of the same texture. In style transfer \cite{gatys2015neural} the goal was to simultaneously match the textural properties of one image (as a representation of artistic style), with the visual content of another, thus allowing the original style to be swapped with that of another image while still preserving its overall semantic content. Both tasks are framed as optimization problems, whereby the statistics of feature maps are used as measures to formulate a loss function between the reference images and the newly generated image. Many authors have extended Gatys' original approach in terms of speed \cite{johnson2016perceptual,ulyanov2016texture}, quality \cite{novak2016improving,ulyanov1607instance}, diversity \cite{ulyanov2017improved} and control \cite{dumoulin2017learned,gatys2016preserving,gatys2017controlling}. For a more thorough review of the historical developments in image style transfer, we refer the reader to Jing et al. \cite{jing2017review} and the individual papers.

To build on existing mechanisms that were developed in the context of the visual domain and carry out the analogous tasks in the audio domain, several exploratory studies have replaced images (in the sense of photographs/pictures) with a time-frequency representation of an audio signal in the form of a spectrogram. 2-dimensional (2D) spectral images have previously been utilized as input representations for CNNs in a range of audio domain tasks  including speech recognition \cite{deng2013deep}, environmental sound classification \cite{salamon2017deep} and music information retrieval \cite{dieleman2014end}. In empirical terms the pairing of spectrograms with CNNs has demonstrated consistently superior pattern learning ability and discriminative power over more traditional machine learning methods such as the hidden Markov model (HMM), and lossy representations like Mel-frequency cepstral coefficients (MFCCs) \cite{ling2015deep}.
 
There have been several preliminary attempts at audio style transfer with varying experimental focus. Ulyanov and Lebedev in a blog post \citep{audio2016style}, outlined several recommendations for audio style transfer including using shallow, random networks instead of the deep, pretrained networks common for the image task. Importantly, while the spectrogram is a 2D representation, in Ulyanov's framework it is still processed as a 1D signal by replacing the colour channel in the image case with the spectrogram's frequency dimension, therefore convolving only across the time dimension. Subsequent work in this field provided a more thorough treatment of the network's effect on the derived statistics. Wyse \cite{wyse2017} looked at network pretraining and input initialization, while Grinstein et al. \citep{grinstein2017audio} focused on the impact of different network architectures on style transfer. In the latter study, the deeper networks, specifically VGG-19 and SoundNet, were contrasted with Ulyanov's shallow-random network and a hand-crafted auditory processing model based upon McDermott and Simoncelli's earlier audio texture work \cite{mcdermott2011sound}. They concluded that the shallow-random network and auditory model showed more promising results than the deeper models. Verma \citep{verma2018neural} in contrast to the others maintained the 2D convolutions and the pretrained network. They found that small convolution kernels are effective for signals with steady-state pitches for the reference style, but provide a limited number of examples which are only presented visually, making interpretation difficult. There was also an attempt to isolate prosodic speech using the style transfer approach by Perez, Proctor and Jain \cite{perez2017style}, although only low-level textural properties of the voice were successfully transferred. 

While all the related works studied certain aspects of the problem, none go into detail on the challenges posed by the nature of sound and how it is represented, especially in relation to what is essentially a vision inspired model in the CNN. Hence, the focus of this paper is not a presentation of state-of-the-art audio style transfer but an analysis of the issues involved in adopting existing style transfer and texture synthesis mechanisms for audio. In paintings, Gatys' style formulation preserves brush strokes, including, to a degree, texture, direction, and colour information, leading up to larger spatial scale motifs (such as the swirls in the sky in Starry Starry Night, see Fig.1 in \cite{gatys2015neural}) as the receptive field grows larger deeper in the network. How these style concepts translate to style in the audio domain is not straightforward and forms part of the analysis here. 

Our main contributions are threefold, and relate to the overall goal of elucidating the possibilities and limitations of existing vision-focused style transfer techniques when applied to the audio domain. Firstly, we distil the issues involved in directly applying CNN-based style transfer to audio. The problems highlighted serve as a basis for the improvements suggested here and elsewhere. Secondly, we provide further insight to possible characterizations of style and content within a style transfer framework given a spectrogram input, as well as its connection to audio texture. Thirdly, we strengthen the link to audio textures by developing a novel use of descriptive statistics taken from deep neural networks for audio texture generation. 

The rest of the paper is organized as follows: in the first half we discuss the theoretical issues that arise when applying the CNN architecture, largely designed to process 2D images, to the domain of audio. The second half documents several experiments that further expound the issues raised in the first half. Here we also show, through an exploration of the relevant architectures, how the style transfer framework may in fact align nicely with the conventional description of audio texture. 

To further illustrate the concepts introduced, audio samples are interspersed throughout the paper. They can be heard, together with additional examples, on the companion website at the following address: \url{http://animatedsound.com/research/2018.02_texturecnn/}      

\section{The style transfer algorithm}
\label{sec:1}
Given a reference content image $c$ and a reference style image $s$, the aim of style transfer is to synthesize a new image $x$ matching the “content” of $c$ and the “style” features drawn from $s$. Gatys et al.  \cite{gatys2015neural} provides precise definitions of “content” and “style” for use in the neural network models that produce images that correlate remarkably well with our intuitive understanding of these terms.

Style is defined in terms of a Gram matrix (Eq. \ref{eq1}) used to compute feature correlations as second-order statistics for a given input. We start by reshaping the tensor representing the set of filter responses $\phi$ extracted in layer $n$ of a CNN to ($BH_n W_n$, $ C_n$) where $B$, $H_n$, $W_n$ and $C_n$ are batch size, height, width and number of distinct filters respectively. The Gram matrix is then obtained by taking the inner product of the $i$th and $j$th feature map at each position $k$, additionally normalizing by $U_n= BH_n W_n$.

\begin{equation} \label{eq1}
G = \frac { 1 } { U _ { n } } \phi ^ { T } \phi = \frac { 1 } { U _ { n } } \sum _ { k } \phi _ { i k } \phi _ { k j }$$
\end{equation}

Given a set of style layers $\mathcal{S}$, the contribution of each layer $n$ to the style loss function is defined as the squared Frobenius norm of the difference between the style representations of the original style image and the generated image.

\begin{equation}
L _ { style } ( s , x ) = \sum _ { n \in \mathcal{S} } \frac { 1 } { C _ { n } } \left\| G \left( \phi _ { n } ( s ) \right) - G \left( \phi _ { n } ( x ) \right) \right\| _ { F } ^ { 2 }
\end{equation}

In contrast, content features are taken directly from the filter responses, yielding the corresponding content loss function between the content representations of the original content image and the generated image in content layers $\mathcal{C}$. 

\begin{equation}
L _ { content } ( c , x ) = \sum _ { m \in \mathcal{C} } \frac { 1 } { C _ { m } } \left\| \phi _ { m } ( c ) - \phi _ { m } ( x ) \right\| _ { 2 } ^ { 2 }
\end{equation}

The total loss used to drive the updating of the hybrid image is a linear combination of the style and content loss functions at the desired layers  $\mathcal{S}$ and  $\mathcal{C}$ of the CNN. Starting from an input image $x$ of random noise (or alternatively a clone of the content image $c$), this objective function is then minimized by gradually changing $x$ based on gradient descent with backpropagation through a CNN with fixed weights to obtain an image of similar textural properties as $s$ mixed with the high-level visual content of $c$. The balance of influence between content and style representations can be controlled by altering the weighting hyperparameters $\alpha$ and $\beta$ respectively. $\alpha$ can be set to 0 to completely remove the contribution of content (or otherwise omit the content operations to save compute time) resulting in pure texture synthesis (in which case the input image is initialized with random noise). 

In practice we also introduce an additional regularization term in the loss to constrain the optimization of $x$ to pixel values between $0$ and $1$. Preliminary experiments showed that imposing a hard bound through clipping the spectrogram values resulted in the saturation of many pixels at $0$ or $1$, negatively impacting the perceived quality of the generated sounds despite a similar approach working reasonably well for images. Instead, the regularization term penalizes values to the extent that they are outside the interval $(0, 1)$, and is weighted by $\gamma$. 

\begin{align}
U &= \max \{ x -  \mathbf {1}, \mathbf {0} \} \\
D &= \max \{ \mathbf { 0 } - x,  \mathbf {0}\}
\end{align} where $\mathbf { 0 }$ and  $\mathbf {1}$ are matrices of 0s and 1s the same size as $x$

\begin{equation}
\begin{aligned}
x &=\argmin_x  L _ { total } ( c , s , x ) \\
&= \argmin_x \alpha L _ { content} + \beta L _ {style} + \gamma \| U + D \| _ { 2 }
\end{aligned}
\end{equation}

\section{Stylization with spectrograms}
\label{sec:2}

\subsection{The spectrogram representation}

For many classification tasks as well as for style transfer, images are represented as a 2D array of pixel values with 3 channels (red, green and blue) for colour or 1 channel for greyscale. Convolutional neural networks have achieved their well-known level of performance on tasks in the visual domain based on this input representation. The traditional 2D-CNN working on visual tasks uses kernels that span a small number of pixels in both spatial dimensions, and completely across the channel dimension. One of the defining characteristics of convolutional networks is that kernels at different spatial locations in each layer share weights. This architecture drastically reduces the number of parameters compared to fully connected layers, making networks more efficient and less prone to overfitting. It also aligns with the intuition that image objects have translational invariance, that is that the appearance of objects does not change with their spatial location. In this sense, the CNN architecture is domain-dependent. 

As a starting point for using existing style-transfer machinery to explore audio style transfer, audio can be represented as a 2D image through a variety of transformations of the time domain signal. One such representation is the spectrogram, which is produced by taking a sequence of overlapping frames from the raw audio sample stream, performing a discrete Fourier transform on each, and then taking the magnitude of the complex number in each frequency bin.

\begin{equation} \label{eq7}
X ( n , k ) = \sum _ { m = 0 } ^ { L - 1 } x ( m ) w ( n - m ) e ^ { - i \frac { 2 \pi k } { N } m }
\end{equation}

Explicitly, the discrete Fourier transform with a sliding window function $w(n)$ of length $L$ is applied to overlapping segments of the signal $x$ as shown in Eq. \ref{eq7}. The overall process is known as the short-time Fourier transform (STFT).   

\begin{figure}[!ht]
  \centering
     \includegraphics[width=0.85\columnwidth]{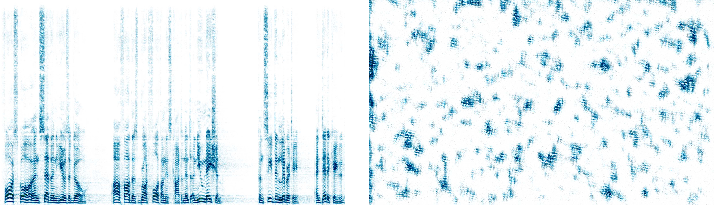}
  \caption{Reference style image $s$ in the form of a spectrogram (left) and synthesized texture $x$ processed using a 2D-CNN model (right), matching the style statistics of $s$. Texture synthesis done with a 2D convolutional kernel results in visual patterns that are similar to those for image style transfer but does not make much perceptual sense in the audio space, as can be heard in the \href{http://animatedsound.com/research/2018.02_texturecnn/\#sec3}{accompanying audio clip}. This study makes several suggestions to better align style transfer to the audio domain. }
  \label{fig1}
\end{figure}
 
 The spectrogram consists of a time-ordered sequence of magnitude vectors, with frequency along the y-axis and time along the x-axis, an example of which is seen in Fig. \ref{fig1}. Where color images have a separate channel for each color, this audio representation has just one for the magnitudes of each time-frequency location. This audio-as-image representation can be used directly with the same architectures that have been developed for image classification and style-transfer, and has been shown to be useful for various purposes in several previous studies \cite{deng2013deep,dieleman2014end,salamon2017deep,audio2016style,wyse2017}.
 
The magnitude spectrogram does not contain phase information. To invert magnitude spectra, reasonable phases for each time-frequency point must be estimated, but there are well-known techniques for doing this that produce reasonably high-quality audio reconstructions \cite{beauregard2015single,griffin1984signal}.

\subsection{Objects, Content, and Style}

What we mean by content and style are necessarily domain dependent. In the visual domain, content is commonly associated with the identity of an object. The identity of an object is invariant to the different appearances it may take. In the audio domain, psychologists and auditory physiologists speak of ``auditory objects'' that are groupings of information across time and frequency which may or may not correspond to physical source \cite{bizley2013and}. The invariance of such objects across different sonic manifestations is a central concept in both domains. However, the relationship between sound and an ``object'' is more tenuous than it is for images. Sounds are more closely tied to events, sonic characteristics are the result of the interaction between multiple objects and their physical characteristics, and sounds often carry less information about the identity of their physical sources than images.

When we consider abstract representations where sense-making is based on the interplay of relationships between patterns rather than on references to real-world objects, then the distinction between content and style becomes even more fluid. The vast majority of images used in the literature on style transfer are not abstract, but are of identifiable worldly objects and scenes. On the other hand, abstract material, like music, occupies a huge part of our sonic experience, and this requires a different understanding of content. The question for this paper is, what are the domain-specific implications for a computational architecture for separating style and content?

There is a conventional use of the term ``style'' in music that, to a first approximation, refers to everything except the sequence of intervals between the pitches of notes. We can play a piece of music slow or fast, on a piano or with a string quartet, without changing the identity of the piece. If we consider content to be auditory objects corresponding to time-frequency structures of pitches, then how appropriate are 2D CNNs and the Gram matrix metric for separating style from content, and how does the representation of audio interact with the computational architecture?

``Pitch'' is a subjective quality we associate with sounds such as notes played on a musical instrument or spoken vowels. To a first approximation, pitched sounds have energy around integer multiples (``harmonics'') of a fundamental. The different components are perceptually grouped and heard as a single source rather than separately. The relative amplitude of the harmonics contributes to the ``timbre'' of the sound that allow us to distinguish between, for example, different instruments producing the same note or different vowels sung at the same pitch. For this reason, we can think of a pitched sound as an auditory ``object''” that has characteristics including pitch, loudness and timbre. There are fundamental differences between the way these auditory objects are represented in spectrograms and the way physical objects are represented in visual images that, if unaccounted for, may impact the efficacy of spectrograms as a viable data representation for CNN-based tasks. 

Objects in images tend to be represented by contiguous pixels, whereas in the audio domain the energy is in general not contiguous in the frequency dimension. CNN kernels learn patterns within continuous receptive fields which may therefore be less useful in the audio domain. Furthermore, in the visual domain, individual image pixels tend to be associated with one and only one object, and transparency is rare. In the audio domain, transparency is the norm and sound emanating from different physical objects commonly overlap in frequency. The hearing brain routinely permits energy from one time-frequency location to multiple objects in a process known as ``duplex perception'' \cite{rand1974dichotic}. 

One of the implications of the distinctions between the 2D auditory and visual representations for neural network modeling appears in how we approach data augmentation. Dilations, shifts, rotations, and mirroring operations are common techniques in the visual domain because we do not consider the operations to alter the class of the objects. These operations are clearly domain-specific since they can completely alter how a sound is perceived when performed in the time and/or frequency domains. Such operations when done in the Fourier space may inadvertently destroy the underlying time domain information, with the recovered audio often losing any discernible structure relating to the original.     

Perhaps the most pertinent issue in using spectrograms in place of images is the inherent asymmetry of the axes. For a linear-frequency spectrogram, translational invariance for a sound only holds true in the x (time) dimension, but not in the y (frequency) dimension. Raising the pitch of a sound shifts not only the fundamental frequency but also results in changes to the spatial extent of its multiplicatively-related harmonics. In other words, linearly moving a section of the frequency spectra along the y-axis would change not only its pitch but also affect its timbre, thereby altering the original characteristics of the sound. In contrast, shifting a sound in time does not change its perceptual identity. Because of the difference between the semantics of space versus time and frequency, as well as the asymmetry of the x and y-axes in the audio representation, we might expect a 2D kernel with a local receptive field (like those used in the natural image domain) that treats both frequency and time in the same way to be problematic when applied to a spectrogram representation (see Fig. \ref{fig1}).

To address this problem, an adjustment can be made to either the way frequency is scaled or the way the model architecture treats the dimensions. One possibility is to use a log-frequency scaling such as the Mel-frequency scale or the constant-Q transform (CQT) that has been empirically shown by Huzaifah \cite{huzaifah2017comparison} to improve classification performance. This is at least suggestive of the influence of input scaling on the pattern learning ability of the CNN, of which translational invariance is a central idea. The CQT is of particular interest as it preserves harmonic structure, keeping the relative positions of the harmonics the same even when the fundamental frequency is changed, preserving some degree of invariance in translation. Sounds that have the same relative harmonic amplitude pattern would then activate the same set of filters independent of any shift in the pitch of the sound.

\begin{figure*}[!ht]
  \centering
     \includegraphics[width=0.90\textwidth]{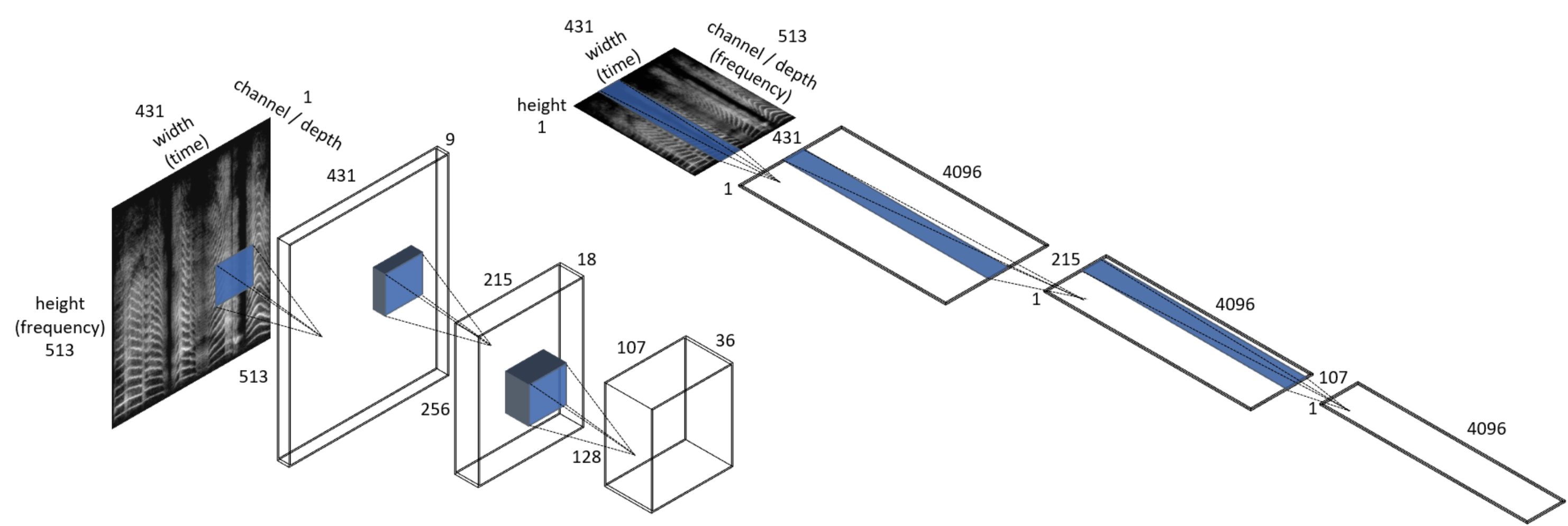}
  \caption{2D (left) and 1D (right) variations of the 3-layer CNN network that was used in this study. Only convolutional layers are shown, with the kernels depicted in blue and the numbers representing the size of each dimension. Note the kernel is shifted only across width in the 1D case. }
  \label{fig2}
\end{figure*}

Alternatively, we can alter the architecture typically used in visual applications. Recent work in the audio domain has turned to the use of 1D-CNN \cite{audio2016style}, where we re-orientate the axes: the spectrogram’s frequency components are moved into the channel dimension, thereby treated analogously to colour in visual applications. Of course, the number of audio channels is thus much greater than the 3 typically used for colour. The time axis remains as the width as in the 2D-CNN which is now treated as the 1-dimensional ``spatial'' input to the network (Fig. \ref{fig2}, right). As in image applications, the convolution kernel still spans the entire channel dimension and has a limited extent of several pixels along the width. We take advantage of this unequal treatment of the dimensions to allow the CNN to learn patterns along the full spectrum that are still local in time. Moreover, by allowing kernel shifts and pooling operations only over the time dimension, this re-orientation of the axes reinforces the idea of kernels capturing translational invariance over time but not over frequency. Equally, one can think of the 1D-CNN arrangement as a 2D convolutional kernel spanning the entire frequency spectrum, valid padding-like, on one edge while shifting the kernel only across the time axis. 

Interestingly, there is also the possibility of transposing the two dimensions, and instead of frequency we can treat time bins as channels. Visually this operation can be pictured as a 90$^{\circ}$ rotation of the spectrogram along an axis perpendicular to its plane in the 1D-CNN in Fig. \ref{fig2}. In this representation, the width over which the kernel slides is now occupied by frequency. This arrangement was not designed to mitigate any issues with the audio representation, and in fact undoes the earlier advantages of the 1D-CNN when it comes to imposing translational invariance only over time (and not frequency), but we use it to explore additional possibilities in texture synthesis. Having the global temporal structure preserved in the channels also allows us to isolate any effects on the frequencies due to style transfer. This second model was therefore adopted for a comparison between linear-scaled STFT and CQT input spectrograms\footnote{See section \ref{sec:74} Frequency scaling.} that have different treatment for frequencies but do not alter things temporally.

\section{Concerning the Gram matrix and audio textures}
\label{sec:3}

The empirical observation that deeper levels in the network hierarchy capture higher level semantic content motivated the use of a content loss based directly on these features in Mahendran \cite{mahendran2015understanding} and later, Gatys \cite{gatys2015neural}. Visualization experiments to help interpret the learned weights have shown that individual neurons, and indeed whole layers, evolve during training to specialize in picking out edges, colour and other patterns, leading up to more complex structures.

Similar attempts have been made to interpret spectrogram inputs to the network. Choi et al. \cite{choi2016explaining} reconstructed an audio signal from ``deconvolved'' \cite{zeiler2014visualizing} spectrograms (actually a visualization of the network activations in a layer obtained through strided convolutions and unpooling) by inverting their magnitude spectra. They found that filters in their trained networks were responding as onset detectors and harmonic frequency selectors in the shallow layers that are analogous to low-level visual features such as vertical and horizontal edge detectors in image recognition. For the 1D-CNN used in this paper, convolution kernels have temporal extent, but just as with image processing networks, the kernels extend along the entirety of the channel depth. As a result, the spectral information occupying this dimension is left mostly intact in feature space. This makes the derived features more easily interpretable especially in the shallower layers.

As for style, Gatys observed that it can be interpreted as a visual texture that is characterised by global homogeneity and regularity but is variable and random at a more local level \cite{davies2008handbook}. As first conjectured by Julesz in 1962 \cite{julesz1962visual}, low order statistics are generally sufficient to parametrically capture these properties. In Gatys' formulation of the style transfer problem, this takes the form of a Gram matrix of feature representations provided by the network.

In the standard orientation of a 1D-CNN with frequency bins as channels, the Gram matrix operation transforms a matrix of features maps indexed by time into a square matrix of correlations, with each correlation value between a pair of feature maps, capturing information on features that tend to be activated together. Further, each entry of the Gram matrix is the sum total of a feature correlation pair across all time. This folding in of the time dimension results in discarding information relating to specific time bins (just as this metric of style discards global spatial structure in visual images). We are thus left with a global summary of the relationship between feature maps abstracted from any localized point of time. We would therefore expect the optimization over style statistics to generally distribute random rearrangements of sonic patterns or events over time. In this case, full-range spectral features over short (kernel-length) time spans are texture elements that are statistically distributed through time to create a texture.

The second variant of the 1D-CNN, referred to in the preceding section, has time bins as channels. In this orientation feature maps would largely retain temporal information while being indexed by frequency. This creates a different kind of texture model where features localized in frequency but with long-term temporal features are statistically distributed across frequency to create a texture.
 
With this insight, parallels can be drawn to the conventional idea of audio textures which are, correspondent to their visual namesakes, consistent at some scale - that scale being some duration in time for the first case and some range in frequency for the second.


Informally, audio textures are sounds that have stable characteristics within an adequately large window of time. Common examples include fire crackling, waves crashing and birds chirping. There have been a variety of approaches to modeling audio textures, including those by Athineos and Ellis \cite{athineos2003sound}, Dubnov et al. \cite{dubnov2002synthesizing}, Hoskinson and Pai \cite{hoskinson2001}, and Schwarz and Schnell \cite{schwarz2010descriptor}. These texture models are all characterized by the distinction between processes at different time scales; one with fast changes, the other static or slowly-varying, and provide representations that can support perceptually similar re-synthesis, temporal extension, and interactive control over audio characteristics.

Neural networks based on style transfer mechanisms from the visual \cite{gatys2015neural,gatys2015texture} and audio \cite{audio2016style} domains offer a new approach to texture modeling. Under this paradigm, texture corresponds to the style definition, while content is trivial given the absence of any large-scale temporal structure. We will explore style transfer architectures further for additional insights and propose several extensions to the canonical system to allow for greater flexibility and a degree of expressive control over the audio texture output.

\section{Experiments}
\label{sec:5}

The rest of the paper documents experiments that probe the issues discussed previously, and show both the working aspects and limitations of CNN-based audio style transfer (Section \ref{sec:6}) and texture synthesis (Section \ref{sec:7}).  

\subsection{Technical details}

To utilize the feature space learned by a pretrained CNN many previous studies have used VGG-19 \citep{simonyan2014very} trained on Imagenet \cite{deng2009imagenet}. Learned filters were initially noted by Gatys to be important to the success of texture generation \cite{gatys2015texture} and by extension style transfer. However, this assessment was revised in a more recent work by Ustyuzhaninov, Brendel, Gatys and Bethge \cite{ustyuzhaninov2016texture} that suggests that random, shallow networks may work just as well for many case conditions. Network weights certainly have some influence on the extracted features and perceptual quality of synthesis as Wyse \cite{wyse2017} reported that audible artefacts were introduced when using an image trained network (i.e. VGG-19) for the audio style transfer task. To further probe this effect in the audio domain, we compare a network with random weights against one trained on the audio-based ESC-50 dataset \cite{piczak2015esc}.

For the audio-trained network, ESC-50 audio clips were pre-processed prior to training as follows: All clips were first sampled at 22050 Hz and clipped/padded to a standard duration of 5 seconds. Spectrograms were generated via short-time Fourier transform (STFT), using a window length of 1024 samples ($\sim$46.4 ms) and hopsize of 256 ($\sim$11.6ms). STFT magnitudes were normalized through a natural log function $\log([1+ |magnitude|)$ and then scaled to (0, 255) to be saved as 8-bit greyscale png images. The samples used for style transfer underwent the same processing without the initial clipping/padding.

For a comparison, a second dataset was prepared containing CQT-scaled spectrograms that were converted from the previous linear-scaled STFT spectrograms using the algorithm adapted from Ellis \cite{ellis2013cqt}, which is invertible, though mildly lossy. Invertibility is required after the style transfer process to recover audio from the new image using a variant of the Griffin-Lim algorithm \cite{beauregard2015single,griffin1984signal}.

\begin{figure}[!h]
  \centering
     \includegraphics[width=0.25\columnwidth]{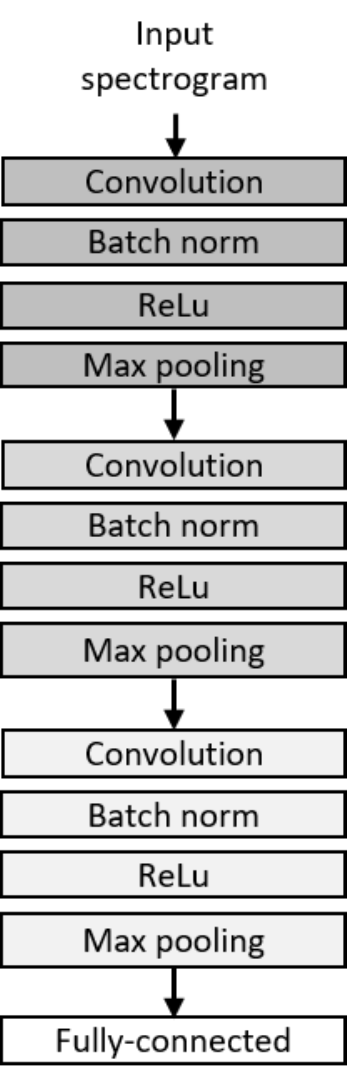}
  \caption{Overview of the network layers used in this study.}
  \label{fig3}
\end{figure}

The network architecture consists of 3 stacks; within each stack a convolutional layer is followed in sequence by a batch normalization layer, ReLu and max pooling (Fig. \ref{fig3}). Our initial investigations and prior work by Ulyanov and Lebedev \cite{audio2016style}, and Wyse \cite{wyse2017} influenced the design of the CNN in an important way. A very large channel depth was required to compensate for the reduction in network parameters for a 1D-CNN, and which generally lead to superior style transfer results. This is corroborated by Gatys in \cite{gatys2015texture}, in which he demonstrated that increasing the number of parameters led to better quality synthesized textures. In each of the convolutional layers we used 4096 channels, much larger than the typical channel depth in image processing networks. Meanwhile, the number of channels in the 2D-CNN were heuristically chosen to roughly match the number of network parameters in the 1D-CNN and hence were much smaller (see Fig. \ref{fig2}). In addition to the 2D-CNN, two variants of the 1D-CNNs were used in the experiments: 1D with frequency bins as channels and 1D with time bins as channels. Unless otherwise stated, all experiments were done with the 1D-CNN with frequency bins as channels and a kernel size of 11. Also for all experiments\footnote{When referring to layers, the nomenclature used in this paper is as follows: $relu1$ refers to the ReLu layer within the 1st stack, $conv3$ refers to the convolutional layer in the 3rd stack etc.}, $\mathcal{C}={relu3}$ and $\mathcal{S}={relu1,relu2}$. Further information on hyperparameters from the classification of ESC-50 and style transfer will be provided in the \hyperref[appendix]{Appendix}. A Pytorch implementation of audio style transfer as used in our experiments is also available on Github\footnote{\url{https://github.com/muhdhuz/Audio_NeuralStyle}}.

\subsection{Audio style transfer}
\label{sec:6}

We attempt audio style transfer with the 1D-CNN described previously, and characterize the output in terms of its perceptual properties and the CNN  architecture. Several technical aspects that effect the result are also investigated.  

\subsubsection{Examples of audio style transfer}
\label{sec:61}

\begin{figure*}[!tbp]
  \centering
     \includegraphics[width=0.75\textwidth]{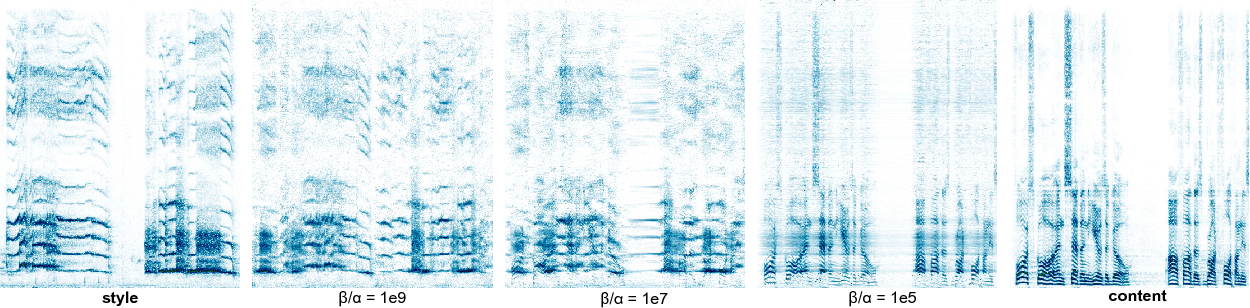}
  \caption{The columns show different relative weightings between style and content given by $\beta/\alpha$, with $s$ on the left and $c$ on the right. Samples can be heard \href{http://animatedsound.com/research/2018.02_texturecnn/\#vary}{here}.}
  \label{fig4}
\end{figure*}

Several examples of audio style transfer were generated, and can be heard on the \href{http://animatedsound.com/research/2018.02_texturecnn/#examples}{companion website}. The content sound (English male speech) was stylized with a variety of different audio clips including Japanese female speech, piano (``River flows in you''), a crowing rooster, and a section of ``Imperial March''.

Though the style examples came from diverse sources (speech, music and environmental sound) and so had very different audio properties, the style transfer algorithm processed them similarly. ``Content'' was observed to mainly preserve sonic objects (pitch), along with their rhythmic structure. Meanwhile ``style'' mostly manifests itself as timbre signatures without clear temporal structure. The mix seemed to be most strongly activated at places where the dynamics of the style and content references coincided. Furthermore, even though the speech does take on the timbre of the style reference, content and style features are not disentangled completely, and the original voice can still be faintly heard in the background. For a contrast, we switched to using piano as the content and speech as the style for synthesis. As before, the interplay between style and content characterizations can be clearly heard. As ``style'', the human voice follows the rhythm of the piano content, without actually forming coherent words. The original piano content was however more prominent than in the previous speech examples and so may sound less convincing as an example of style transfer to the reader.

While audio style transfer has been shown here to succeed on a wide variety of sources, it should be noted that individual tuning was required to obtain a subjectively ``good'' balance of content and audio features. It is perhaps the case that style and content references with similar feature activations would lead to smoother optimization and a better meshing during style transfer, but this would limit workable content/style combinations in a way that is not necessary in visual style transfer. 

\subsubsection{Varying influence of style and content}
\label{sec:62}

The preceding examples were synthesized with careful tuning of the weighting hyperparameters to obtain perceptually satisfying examples of style transfer. To further illustrate the \href{http://animatedsound.com/research/2018.02_texturecnn/\#vary}{effect of mixing different amounts of style and content features}, we progressively varied the ratio of weighting  hyperparameters $\alpha$ and $\beta$ as shown in Fig. \ref{fig4}.The content target here was speech, while the style target was a crowing rooster. 

For $\beta/\alpha = 1e5$, the original speech can be generally heard clearly in the presence of some minor distortions and audio artefacts. From the spectrogram it becomes clear that the distortions originate from the faint smearing of the dominant frequencies in the reference style across time. The mixing of the dominant frequencies for both the speech and rooster sounds becomes more apparent when $\beta/\alpha = 1e7$. At this ratio, the original speech can still be heard albeit with less prominence, while it starts to take on the timbre of the rooster. Further increasing the influence of style at $\beta/\alpha = 1e9$ results in the output completely taking on the timbre of the rooster. It is interesting to note that the rhythm of the speech at this ratio is still retained. 

Using the characterization of style as timbre and content as pitch and rhythm as before, we conclude that the style transfer at $\beta/\alpha = 1e9$ generates the best result among the examples here, although more fine tuning (as was the case in the preceding section) would arguably generate better results.

\subsubsection{Network weights and input initialization}
\label{sec:63}

\begin{figure}[!b]
  \centering
     \includegraphics[width=0.60\columnwidth]{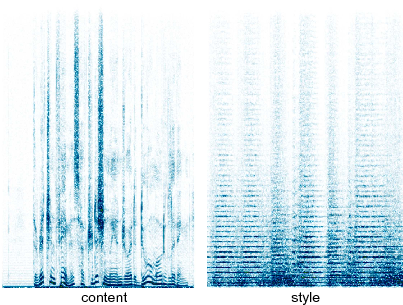}
  \caption{Reference content image $c$ (left) and reference style image $s$ (right). Listen to them \href{http://animatedsound.com/research/2018.02_texturecnn/\#weightsninput\_ref}{here}.}
  \label{fig5}
\end{figure}

\begin{figure}[!htbp]
  \centering
     \includegraphics[width=0.60\columnwidth]{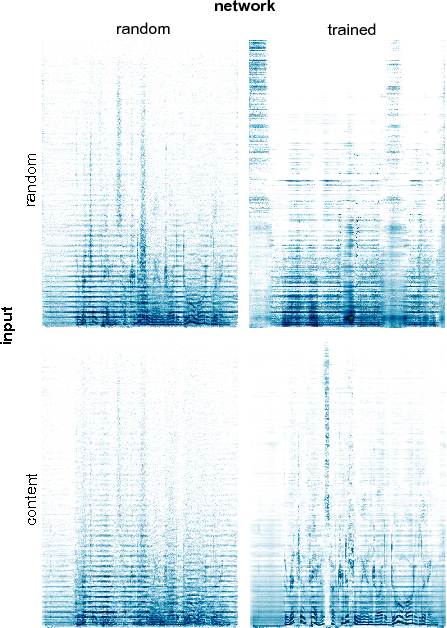}
  \caption{\href{http://animatedsound.com/research/2018.02_texturecnn/\#weightsninput\_results}{Synthesized hybrid image $x$ with different input and weight conditions}. Clockwise from top left: (1) With random weights and initialization, the speech is clear with the style timbre sitting slightly more recessed in the background (2) It was still fairly problematic to get good results for trained weights and a random input. Hints of the associated style and content timbres could be picked up and seem well integrated but are very noisy. The output also loses any clear long term temporal structure (3) Trained weights and input content generated a sample sound fairly similar to (1) with a greater presence of noise and artefacts (4) Random weights plus input content seem to result in the most integrated features of both style and content.}
  \label{fig6}
\end{figure}

Stylized spectrograms were generated with different input conditions and network weights as shown in Fig. \ref{fig6}. The synthesized image $x$ was either initialized as random noise or a clone of the content reference spectrogram $c$. For stylization with a trained network, the 1D-CNN was pretrained for the classification of the ESC-50 dataset for 30 epochs. \href{http://animatedsound.com/research/2018.02_texturecnn/\#weightsninput\_ref}{The content target was speech, while the style target was ``Imperial March''} (Fig. \ref{fig5}).

\begin{figure*}[!tbp]
  \centering
     \includegraphics[width=0.90\textwidth]{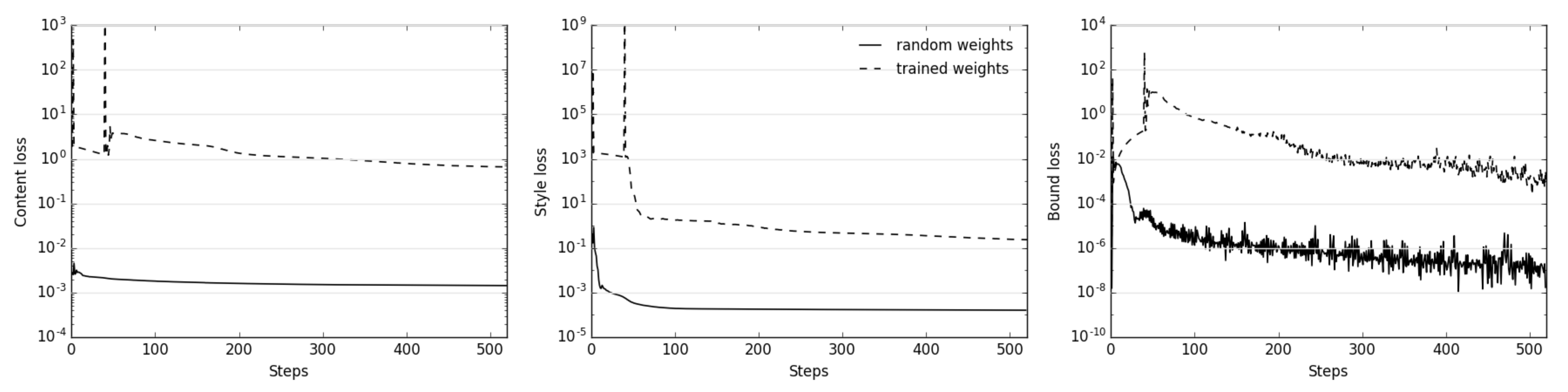}
  \caption{Losses from the optimization process comparing the difference between random weights depicted as solid lines and trained weights depicted as dashed lines. Tuning the hyperparameters lead to generally better convergence for both random and trained weights, and more integrated style and content features. Despite this, losses were still very large for trained wights resulting in a much slower convergence.}
  \label{fig7}
\end{figure*}

Among the \href{http://animatedsound.com/research/2018.02_texturecnn/\#weightsninput\_results}{four variations tested}, the combination of random network weights and content initialization brought about the most stable style transfer. Trained weights generated much higher content and style losses during the optimization process as compared to random weights, regardless of whether the sounds used for style transfer were previously exposed to the network as part of the ESC-50 dataset. A possible reason for this observation may be the more powerful discriminative power of a network pretrained for classification, resulting in a high Frobenius norm between features derived from different inputs. The high initial losses with trained weights had a negative impact on the optimization, leading it to fall into a local minimum (Fig. \ref{fig7}) and an ensuing noisy-sounding synthesis from the network (Fig. \ref{fig6}, right). Indeed, a more aggressive regularization penalty in the form of $\gamma$ had to be used for the optimization to proceed smoothly. However, the optimization took much longer to converge.

As expected, initializing with a clone of the content reference spectrogram resulted in very strong content features in the output, with the style features normally less pronounced (Fig. \ref{fig6}, bottom). In contrast, starting from a noisy image yielded more mixed results. In Fig. \ref{fig6} (top left), the output sounds more like a mix of content and style rather than stylized content. With pretraining (Fig. \ref{fig6}, top right), the output sounds heavily distorted on both the content and style.

Interestingly, with some hyperparameter tuning ($\alpha$, $\beta$ and $\gamma$), all four variants, at least by visual inspection of the spectrograms, look to be plausible results of style transfer. All outputs generally display the style frequencies overlaying that of the content, while still retaining the onset markers of the content sound. Nevertheless, not all of them are of the same sonic quality, and listening to the corresponding audio clips show a significant difference between them. Using a random network with content initialization consistently rendered more integrated style and content features than the others. The observation that the expectation of style transfer in images and audio do not always align may be indicative of a greater sensitivity in the audio domain to what constitutes ``successful'' style transfer in comparison to vision, suggesting that the manifold of such examples exists in a narrower range of data space than for visual data. 

Overall, our results show that trained weights are not essential for successful style transfer, and the particular combination of hyperparameters had a greater impact on the output, therefore validating Gatys' revised conclusion in \cite{ustyuzhaninov2016texture}. On the other hand, random weights seemed to be more tolerant to changes in the hyperparameters as a wider range and more combinations of hyperparameters led to successful style transfer. Samples generated from random weights were also generally found to be cleaner in comparison to trained weights that slightly coloured the output with audio artefacts that do not appear to be from the reference style or content. This final result is in accordance with Ulyanov \cite{audio2016style} and Grinstein et al. \cite{grinstein2017audio} who observed that random networks worked at least as well or better than pretrained networks for audio style transfer, suggesting that this behaviour may be even more pronounced for the audio domain. 

\subsection{Audio texture synthesis}
\label{sec:7}

The previous sections demonstrate how a degree of style transfer is possible with the CNN-based framework although limitations exist in obtaining expressions of audio content and style as lucid as those seen in image work. However, if we leave the task of disentangling and recombining the divergent aspects of content and style behind, and consider only a single perceptual factor (i.e. style as texture), we find that the Gram matrix description of audio texture is a fairly representative summary statistic when applied to either one of frequency or time. In this section, several aspects of audio texture synthesis by the style transfer schema are discussed further. 

\subsubsection{Infinite textures}
\label{sec:71}

Important to the versatility of the technique for texture synthesis is the possibility of continuous generation i.e. infinite synthesis. This is achieved by appreciating the fact that the entries of the Gram matrix divided by $U_n$ (Eq. \ref{eq1}) represent a mean value in time for the normally oriented spectrogram images (the aforementioned abstraction from time). The time axis given by the width of the spectrogram can therefore be arbitrarily extended while still keeping the number of terms in $G(\phi_n (s))$ and $G(\phi_n (x))$ consistent. Consequently, the generated image $x$ need not match the width of the reference style image $s$. This implies that audio textures can be synthesized ad infinitum (or conversely shortened). 

\href{http://animatedsound.com/research/2018.02_texturecnn/\#infinitetex}{Several samples} were synthesized up to 3 times the original input duration to demonstrate this effect. The same style statistics hold for the full duration of $x$, so textural elements are uniquely distributed throughout the specified output length and are not merely repeats of the original duration.

\subsubsection{Multi-texture synthesis}
\label{sec:72}

Textures from multiple sources can be blended together by utilizing the batch dimension of the input, a different mechanism from what has been proposed for multi-style transfer in the literature so far. Previous approaches have included using an aggregated style loss function with contributions from multiple networks trained on distinctive styles \cite{cuimulti}; or combining two styles of distinct spatial structure (e.g. course and fine brushstrokes) by treating one as an input image and the other as the reference style \cite{gatys2017controlling} afterwards using the new blended output as the reference style for subsequent style transfer. Dumoulin et al. \cite{dumoulin2017learned} introduced a ``conditional instance normalization'' network layer that can not only learn multiple styles but also interpolate between styles by tuning certain hyperparameters. 
  
\begin{figure}[!htbp]
  \centering
     \includegraphics[width=0.95\columnwidth]{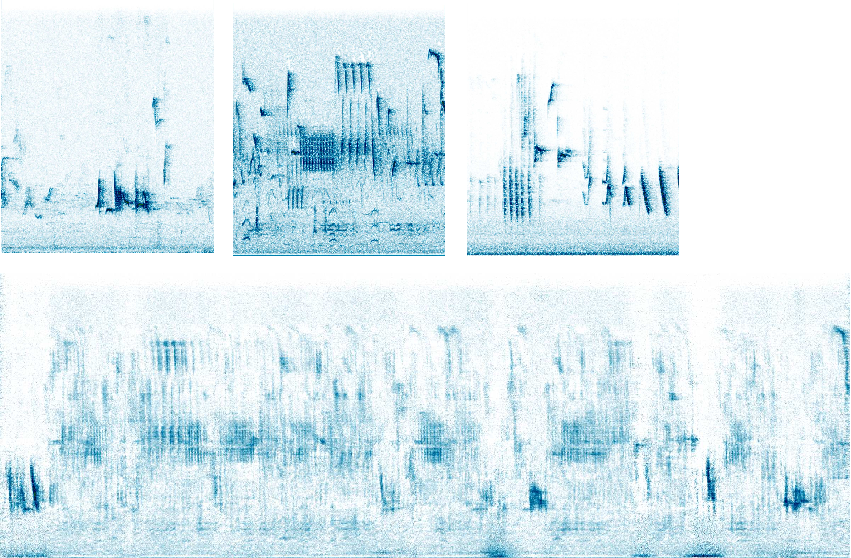}
  \caption{The extended \href{http://animatedsound.com/research/2018.02_texturecnn/\#multitex1}{multi-texture bird sound sample that was synthesized (bottom) from its 3 constituent clips (top)}. Prominent spectral structures from the component sounds can be seen in the output spectrogram although there are overlaps with other frequencies.}
  \label{fig8}
\end{figure}

\begin{figure}[!htbp]
  \centering
     \includegraphics[width=0.75\columnwidth]{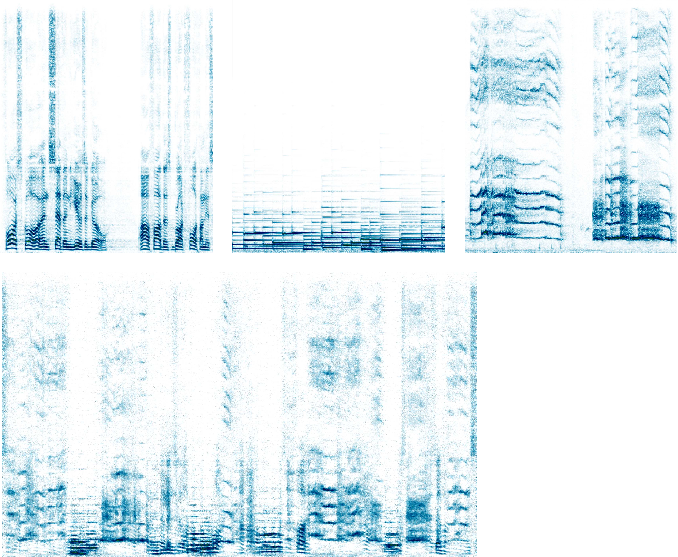}
  \caption{The\href{http://animatedsound.com/research/2018.02_texturecnn/\#multitex2}{ multi-texture output (bottom) from 3 distinct sounds (top)} that may be described as uncharacteristic for textures since they are fairly dynamic in the long term. In contrast to Fig. 6a, the synthesis is not as integrated largely because the constituent sounds have very different spectral properties, resulting in stark transitions between them.}
  \label{fig9}
\end{figure}

A single style image $s$ is represented by a 4-dimensional tensor $(B, C, H, W)$, or $(1, 513, 1, 431)$ substituting the actual numerical values used in our experiments for a 1D-CNN. In our approach to texture mixing, a set of $\Sigma$ style images is fed into the network as $(\Sigma, 513, 1, 431)$. Convolution is done separately on each batch member in the forward pass, resulting in distinct feature maps for each style. Next the batch dimension is flattened and concatenated into the 1st dimension as part of reshaping $\phi$ to $(BHW,C)$ prior to calculating the Gram matrix. As a result, derivative style features are correlated not only within a style image but across multiple styles, producing a convincing amalgamation of textures from several sources.
In this way, more complex environmental sounds can be built up from its components, although we observe that using sources that are similar to each other creates a more cogent overall texture. To demonstrate this, a \href{http://animatedsound.com/research/2018.02_texturecnn/\#multitex1}{multi-texture output was generated from a combination of 3 distinct bird chirping sounds} of 5 seconds each taken from the ESC-50 dataset (Fig. \ref{fig8}). The synthesized output was extended to a length of 20 seconds using the technique outlined in Section \ref{sec:71}. As per the description for texture, chirps from all 3 clips do not merely appear periodically but occur randomly and even overlap at times throughout the output. For a better intuition of the meshing of multi-textures we also \href{http://animatedsound.com/research/2018.02_texturecnn/\#multitex2}{generated a clip made up of 3 very contrasting sounds from speech, piano, and a crowing rooster} (Fig. \ref{fig9}). For this example, there is usually one dominant texture (drawn from just one of the original images) at any given time. It is likely that in this case very few correlations in the Gram matrix are found between sounds from the different sources, leading to the lack of hybrid combinations, but the example still provides insight into the time scales and process of reconstruction.

The technique is not without its drawbacks. Since we use the entirety of the reference textures, unwanted ``background'' sounds from the sources also appear in the reconstructed output, such as the low rumbling noise from one of the \href{http://animatedsound.com/research/2018.02_texturecnn/\#chirp3}{chirp sounds} that is intermittently heard in the multi-texture bird sound example. The naive remedy to this is to essentially make sure the entirety of the reference clip is desirable, although more fine-tune control of certain aspects of the feature space similar to those currently being developed in image style transfer constitutes future work.

\subsubsection{Controlling time and frequency ranges of textural elements}
\label{sec:73}

\begin{figure*}[!tbp]
  \centering
     \includegraphics[width=0.90\textwidth]{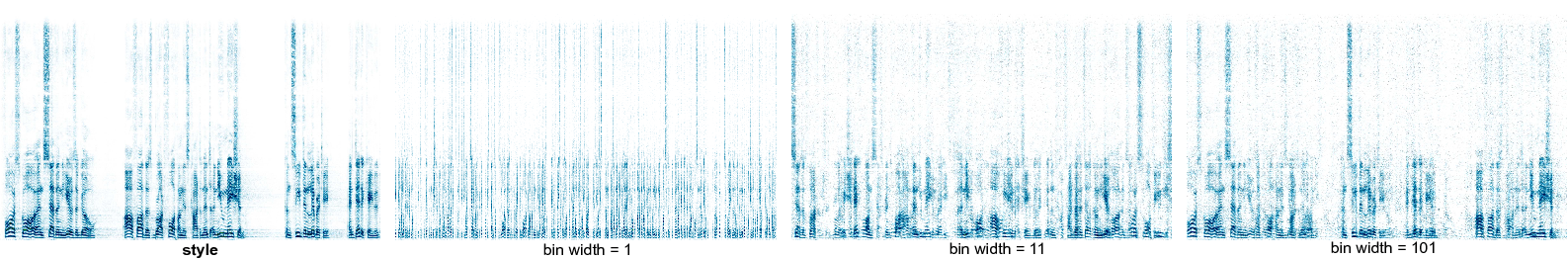}
  \caption{The corresponding spectrogram outputs for an increasing kernel width over time bins and the style reference on the left. The effect of the varying kernel width is apparent from the spectrograms as bigger slices of the original signal is captured in time with a larger receptive field. Samples are found \href{http://animatedsound.com/research/2018.02_texturecnn/\#timerange\_speech}{here}.}
  \label{fig10}
\end{figure*}

The range in time or frequency captured by each textural element can be altered by varying the kernel width hyperparameter in the 1D-CNN, which in turn changes the effective receptive field of neural units in the CNN. This offers us a simple parametric measure to change the desired output.

The effect is most obvious for a 1D-CNN with frequency as channels, in which we control the duration of individual textural elements through the width of the kernel. Several \href{http://animatedsound.com/research/2018.02_texturecnn/\#timerange\_speech}{clips derived from speech} were synthesized with kernel widths ranging from 1 , 11, to 101 bins (Fig. \ref{fig10}). The shortest width of 1 only picks up parts of single words, reminiscent of phonemes, and produces babble. As we lengthen the time window, whole words and eventually phrases are captured. \href{http://animatedsound.com/research/2018.02_texturecnn/\#timerange\_piano}{Further examples} of the increased temporal receptive field can be found in the piano samples. The piano attack is also audibly not as crisp as the original clip due to a degree of temporal blur induced by convolution and pooling. 

In the \href{http://animatedsound.com/research/2018.02_texturecnn/\#freqrange_speechstft}{alternative orientation} with time bins as channels, the kernel width has control over the range in frequency. Using a short width (bin width = 1), local frequency patterns are captured and translated to the entire spectrum, resulting in a sharp, high-pitched version of the original voice, as one would get with sounds containing many extra high frequency harmonics. Nevertheless, the onset of each word is evident enough to conclude that long term time structure is preserved. Using progressively longer kernels (bin width = 11, 101) leads to the output voice becoming closer and closer to its original form as illustrated in Fig. \ref{fig11} (left). This same trend appears for the \href{http://animatedsound.com/research/2018.02_texturecnn/\#freqrange_pianostft}{piano samples}, where the translated high frequency harmonics results in the piano timbre taking on a bell-like quality (bin width = 1, 11) that gets closer to the original with a longer kernel (bin width = 101).

\subsubsection{Frequency scaling}
\label{sec:74}

\begin{figure}[!htbp]
  \centering
     \includegraphics[width=0.95\columnwidth]{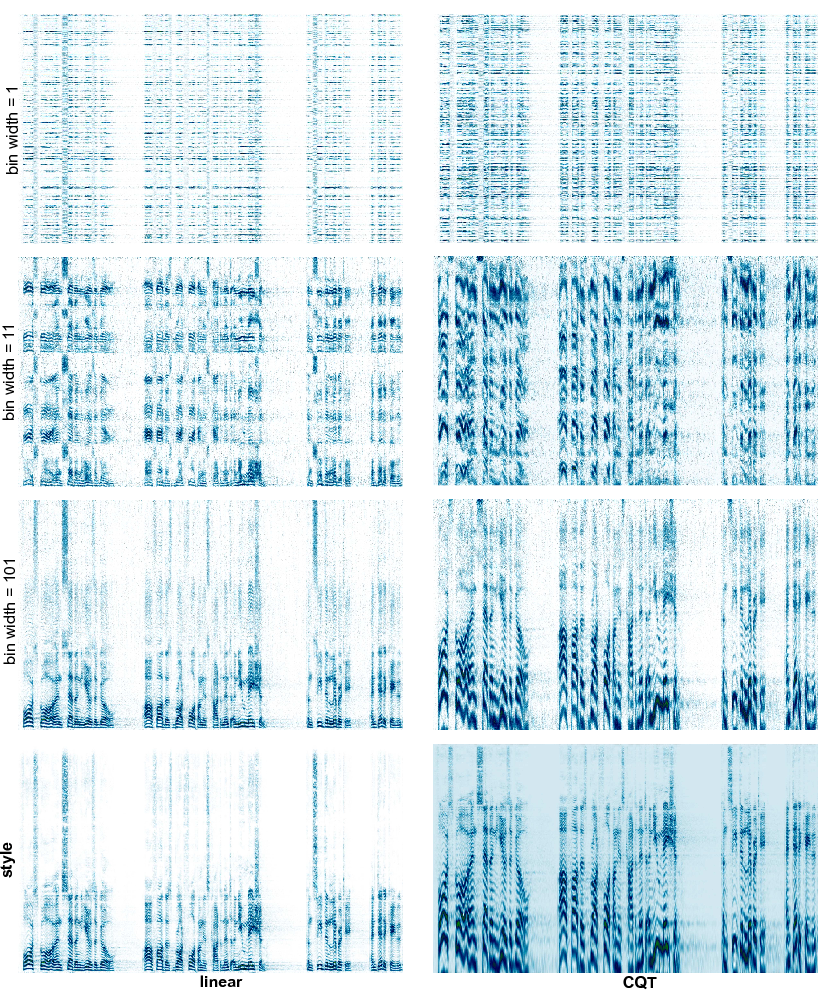}
  \caption{\href{http://animatedsound.com/research/2018.02_texturecnn/\#fig11}{Synthesized textures} generated from linearly-scaled STFT spectrograms (left column) and CQT spectrograms (right column). The style references are shown at the bottom. Gradually lengthening the kernel width (1 , 11, 101 bins) in the re-oriented 1D-CNN results in capturing more of the frequency spectrum within the kernel for both the linear-STFT and CQT spectrograms, although there are subtle differences in the corresponding audio reconstructions.}
  \label{fig11}
\end{figure}

The linearly-scaled STFT spectrograms in the previous experiment were replaced with CQT spectrograms while keeping the kernel width variation to the same 1, 11 and 101 bins for \href{http://animatedsound.com/research/2018.02_texturecnn/\#freqrange_speechcqt}{speech} (Fig. \ref{fig11}, right) and for \href{http://animatedsound.com/research/2018.02_texturecnn/\#freqrange_pianocqt}{piano}. Specifically, we want to test the hypothesis discussed above that the CQT is better suited for translational invariance in the audio frequency dimension. Unlike linear scaling, the CQT preserves harmonic structure as pitch shifts. Conversely, any shift along the frequency dimension in a CQT spectrogram image only affects the pitch and leaves the timbre, or more precisely, the relative amplitudes of harmonically-related components intact. For this experiment the 1D-CNN with time bins as channels was used since it preserves temporal structure thus isolating any effects due to the shifting shared-weight kernels to just the frequencies, allowing for a clearer analysis.

Similar to the linear spectrograms, extending the kernels resulted in the capture of a bigger portion of the spectrum. The two transformations however modified the sounds in distinct ways. One reason is simply that the CQT samples lower frequencies more densely than the linear scaling, and higher frequencies less densely.

Learning localized frequency patterns over the linear-STFT resulted in a single synthesized voice that sounded as if it contained many additional harmonics. In contrast, the CQT generated something that seemingly contained multiple voices at different pitches. An explanation of this phenomenon may lie in the different effect each transformation has on harmonic relationships and how we perceive these relationships. Within the kernel’s receptive field, the CQT maintains harmonic structure and thus sounds are perceptually grouped to sound as part of the same source. Multiple instances of these harmonic groups appear in the synthesis leading to the presence of multiple layered voices. Meanwhile, the linear translation does not preserve the multiplicative relationship between the components and thus the harmonic relationship of the sound breaks. As a result, voices and instruments are heard as a plethora of different individual localized frequency sources rather than as a smaller number of rich timbral sources. In that sense, the CQT achieves translational invariance to a certain degree within the receptive field of the convolution kernel, although timbre as a whole is not preserved because only parts of the full spectrum are being translated. We note, however, that the Gram matrix represents correlations across filters at the same location. In the case of the rotated spectrogram orientation, these are temporally-extended filters located at the same local frequency region. The distributed nature of pitched sound objects that have energy at multiples of a fundamental frequency means that the Gram matrix ``style'' measure is less sensitive to patterns in audio representations than to patterns in visual images, irrespective of the frequency scaling. 

\section{Discussion}
\label{sec:8} 

In this paper we took a critical look at some of the problems with the canonical style transfer mechanism, that relies on image statistics extracted by a CNN, when adapting it for style transfer in the audio domain. Some key issues that were raised include the asymmetry of the time and frequency dimensions in a spectrogram representation, the non-linearity and lack of translational invariance of the frequency spectrum, and the loose definitions of style and content in audio. This served as motivation to explore how style transfer may work for audio, leading to the implementation of several tweaks based around the use of a 1D-CNN similar to the one introduced by Ulyanov and Lebedev \cite{audio2016style}. 

While changes were made to the model architecture and input representations to better align the method with our perceptual understanding of audio, we retained the core idea of using descriptive statistics derived from CNN feature representations to embody notions of content and style. Moreover, how content is  distinguished from style -- direct feature activations versus feature correlations via the Gram matrix (or equivalently first and second-order treatment of network features) -- was similarly adopted unchanged. A major goal was to therefore relate what these metrics mean in terms of audio, and whether they agree with some of our preconceived sentiments of what these concepts represent.

For audio, everyday notions of ``style'' and ``content'' are even harder to articulate than for images. In addition to the technical issues examined experimentally, they tend to be dependent on context and level of abstraction. For example, in music, at a very high level of abstraction, style may be a term for genre and content may refer to the lyrics. At a lower level, style can refer to instrumentational timbres and content to the actual notes and rhythm. On the other hand, in speech, style may relate to aspects of prosody such as intonation, accent, timing patterns, or emotion while content relates to the phonemes building up to the words being spoken.
   
Gatys et al. \cite{gatys2015neural} found that first and second-order treatment of network features relate remarkably well with what we visually perceive as content and style. In a few words, ``style'' is described as a multi-scale, spatially invariant product of intra-patch statistics that emerges as a texture, and ``content'' is the high-level structure, encompassing its semantics and global arrangement. In the audio realm, we have shown through various examples that this spatial structure makes the most perceptual sense when extended to the dimensions of frequency or time separately. 

Perceptually we found that with the style transfer architecture ``content'' strongly captures the global rhythmic structure and pitch while ``style'' characterizes more of the timbral characteristics. Style transfer for audio in this context can hence be construed as a kind of timbre transfer. As a whole however, there is less obvious disentanglement between style and content in comparison to image style transfer, and when their features are combined, they are not as seamless. Consequently, results heard in the audio clips are generally less satisfying and aesthetically appealing than those produced in the image domain.

Despite this, an analysis of the Gram matrix of feature correlations reveals that this particular aspect of the style transfer formulation bears many similarities to the concept of audio textures. Generating new textures of the same type as the reference in this paradigm is a case of rearranging, at a certain scale (largely dictated by the kernel width), either the feature maps of the frequency spectrum or that of temporal information, while leaving the other dimension relatively intact. These ideas were demonstrated with numerous synthesized examples of audio texture, including ones of multi-source textures and the constant-Q representation of audio.  

It is evident that the approach borrowed from the vision domain is far from a complete solution when it comes to imposing a given style on a section of audio, even if we narrow the scope of ``style'' to timbre. Our intuitions about audio style and content are not well captured by the CNN/Gram matrix computational architecture and how it works with time-frequency representations of audio. Furthermore, the finding that random networks were better than trained networks at the audio style transfer task even puts to question the central role of feature representation in a CNN-based style transfer framework. Future work needs to instead develop a better understanding of what constitutes ``style'' and ``content'' in audio. That insight can then be used to further investigate audio representations and model architectures that are a better fit and more specifically suited for audio generative tasks than the visual domain-specific CNN style transfer networks.


%
%


\bibliographystyle{spbasic}      
\bibliography{refs}   

%
%

\section*{Appendix}
\label{appendix}

All models implemented in Pytorch.

\textbf{Model parameters}:
\begin{itemize}
\item Max pooling: 1x2, stride 1x2 (1D-CNN) / 2x2, stride 2x2 (2D-CNN)
\item Padding: ``Same'' padding used on all convolutional layers
\item Batchnorm: Default Pytorch parameters used on all batch normalization layers
\end{itemize}

\textbf{Classification hyperparameters}:
\begin{itemize}
\item Dataset: ESC-50 (fold 1 used for validation, fold 2-5 used for training)
\item Optimizer: Adam (lr = 1e-3, $\beta$1 = 0.9, $\beta$2 = 0.999, eps = 1e-8)
\item Loss criterion: Cross Entropy Loss
\item Batch size: 20
\item Training epochs: 30 
\end{itemize}

\textbf{Style transfer hyperparameters}:
\begin{itemize}
\item Optimizer: L-BFGS (with default Pytorch parameters)
\item No. of iterations: 500
\end{itemize}

\textbf{Section \ref{sec:62}}:
\begin{itemize}
\item Random network/content input: $\alpha$ = 1, $\beta$ = 1e5/1e7/1e9, $\gamma$ = 1e-3
\end{itemize}

\textbf{Section \ref{sec:63}}:
\begin{itemize}
\item Pretrained network/random input: $\alpha$ = 1, $\beta$ = 1e3, $\gamma$ = 1
\item Random network/random input: $\alpha$ = 1e1, $\beta$ = 1e7, $\gamma$ = 1e-3
\item Pretrained network/content input: $\alpha$ = 1, $\beta$ = 1e5, $\gamma$ = 1e1
\item Random network/content input: $\alpha$ = 1, $\beta$ = 1e8, $\gamma$ = 1e-3
\end{itemize}

\textbf{Section \ref{sec:7}}:
\begin{itemize}
\item Random network/random input: $\alpha$ = 0, $\beta$ = 1e9, $\gamma$ = 1e-3
\end{itemize}

\end{document}